\documentclass[osajnl,twocolumn,superscriptaddress,showpacs,10pt]{revtex4-1} 
\usepackage{lineno}

\usepackage{amsmath,amssymb,graphicx}
\usepackage{soul}
\usepackage{afterpage}

\begin{document}
\title{
Self-calibrating Fiber Spectrometer for the Measurement of Broadband 
Downconverted Photon Pairs
}

\author{E. Y. Zhu,$^{*1}$  
            C. Corbari,$^{2}$ 
            P. G. Kazansky,$^{2}$
              L. Qian$^{\dagger}$             
            }
\address{
Dept. of Electrical and Computer Engineering, University of Toronto, 10 King's College Rd., Toronto, ON M5S 3G4, Canada\\
$^2$Optoelectronics Research Centre, University of Southampton, SO17 1BJ, United Kingdom \\
$^*$Corresponding author: eric.zhu@utoronto.ca\\
$^\dagger$Corresponding author: l.qian@utoronto.ca
}

\begin{abstract}
We experimentally demonstrate a simple method to measure the 
biphoton joint spectrum
by mapping the spectral information onto the temporal domain using a dispersive medium.  
Various top-hat spectral filters are used to limit the spectral (and hence, temporal) extent of the 
broadband downconversion photons measured.  The sharp edges of the spectral filters are 
utilized as spectral markers for dispersion characterization of the dispersive medium. This method 
allows dispersion characterization and joint spectral measurement to be completed 
simultaneously.
The 
joint spectrum  (which extends beyond 100 nm, centered about $1.5\ \mu{\text{m}}$) 
of the type-II downconverted photon pairs generated from a poled optical fiber is obtained with this method.  
\end{abstract}


\maketitle 

\section{Introduction}
\label{sec:intro}

Understanding the spectral correlations present in a correlated photon pair source is extremely important in determining whether that source is suitable for various quantum interference and quantum information processing schemes.  
For instance, a photon pair whose signal and idler photons are spectrally-uncorrelated may work well as a heralded source \cite{humble2008effects,Bruno:14,EvansHumble2010},  but be unsuitable for applications where spectral entanglement is required, such as quantum-enhanced optical coherence tomography \cite{abouraddy2002quantum}, or high-dimensional quantum key distribution \cite{nunn2013large,BunandarHDQKD2015}.  

However, traditional measurements of the joint spectrum of a biphoton require the use of expensive and lossy spectrometers, involving long integration times to produce 
significant coincidence counts.  The integration times are  increased when the resolution required is finer.  
Recent progress in so-called `stimulated' emission tomography \cite{StimETomo_Sipe2013} 
has shown that the joint spectrum can be predicted accurately by performing 
classical nonlinear frequency-mixing.  
In practice \cite{fang2014fast}, though, 
this form of tomography requires complete knowledge of and access to the nonlinear 
generation process (which will not be the case if the source is embedded or packaged in a 
commercial system).

Another way to approach the problem is to map the spectral information of the 
JSI onto the temporal domain, using a dispersive medium, so that the spectral information is now encoded in the arrival times of the signal and idler photons at 
their respective detectors.  When the dispersive medium is a long spool of fiber, this is called the fiber spectrometer method \cite{SilberHorn2009}.  
%
%
However, the fiber spectrometer requires accurate dispersion calibration that is specific to the actual fiber spool in use. In \cite{SilberHorn2009,GerrittsHOM2014}, for example, the fiber dispersion is calibrated beforehand with a short-pulse tunable laser.
This must be done before every experiment, as
the temperature changes in Corning SMF-28 fiber, say, can result in a change of group delay by 
40 ps/km/K \cite{bousonville2009velocity}, or 4 ns over 20 km of fiber for a 5 K temperature 
variation.  

In  this  work,  we  demonstrate  experimentally  that  such  \emph{a 
priori} calibration  is  unnecessary,  especially  if  we  wish  to 
measure  a  biphoton  joint  spectrum  that  is  broadband.   By 
using multiple frequency-conjugate filters that have sharp cut-on and cut-off wavelengths, the                                                                                                                                                                                                                                                                   broadband spectrum we obtain a new way to elucidate 
the dispersion of the fiber spool.  This is because the sharp edges of the filter can now 
act as spectral markers, allowing us to simultaneously gauge the fiber spool's dispersion 
while obtaining the joint spectrum of the biphoton's JSI.  
Additionally, the use of the filters limits the temporal extent of the downconverted biphotons, 
so that a relatively high-repetition rate pump (Ti:Sapphire laser, 81.6 MHz, 12.25 ns period) 
can be used without the downconverted biphotons spanning more than one period, which could cause complications experimentally.


The breakdown of this paper is as flows.  
In Section 2, the idea of the fiber spectrometer will be explained mathematically.  
Using downconverted photon pairs generated from a poled optical fiber 
\cite{bonfrate1999parametric,PhanHuy:07,zhueyiPRL2012, zhueyiOL2013} and various top-hat sharp-edge filters, we will demonstrate our  method to measure the dispersion of the fiber spool in Section 3.
Finally, in Section 4, the fiber spectrometer method will be used to measure the downconversion spectrum of the poled optical fiber.    
Conveniently, the downconverted photon pairs generated have wavelengths symmetric about $\sim$1550 nm.  

\afterpage{\clearpage}

\section{The fiber spectrometer}

Consider a biphoton generated from {SPDC}.  Both photons are created
at the same time and will also exit the nonlinear medium simultaneously; in this case, `simultaneous' shall mean within the detector response time (120 ps full width at half maximum [FWHM]). 
 Let us describe this biphoton as a function of wavelength:
 \begin{equation}\label{eq:biphotonN}
 	\bar{N}_{ph} = \int d\lambda_s d\lambda_i\  \tilde{n}(\lambda_s, \lambda_i) ,
 \end{equation} 
\noindent
where $\bar{N}_{ph}$ is the average number of photon pairs in the wavepacket, and $\tilde{n}(\lambda_s, \lambda_i) $ is the 
spectral brightness (photon pairs per unit wavelength of signal and idler, [pairs/nm$^2$]).

Now, let this biphoton propagate through a fiber spool of length $L$ whose dispersion is given by the $D$-parameter \cite{agrawal2005lightwave}:
\begin{equation}
	D(\lambda) = -\frac{2\pi c}{\lambda^2} \frac{d^2 \beta}{d \omega^2},
\end{equation}
where $c$ is the speed of light in vacuum and $\beta$ is the fiber propagation constant.   
The fiber is single-mode at the wavelengths of the downconverted photon pairs.    
We now exploit the fact that single photons travel at the group velocity \cite{SteinbergSinglePhoton} to develop the fiber spectrometer idea further.

%
The spectral information of the biphoton will then be mapped onto the temporal domain:
\begin{equation}\label{eq:biphotTime}
	\bar{N}_{ph} = \int dt_s\  dt_i\ 
   {n}(t_s(\lambda_s), t_i(\lambda_i)),
\end{equation}
where we have explicitly stated that the arrival time $t_s$ ($t_i$) of the signal (idler) is  a function of its wavelength $\lambda_s$ ($\lambda_i$).  It is a simple matter, then, to relate ${n}$ (Eq. \ref{eq:biphotTime}) to $\tilde{n}$ (Eq. \ref{eq:biphotonN}).  Noting that the arrival time is defined as 
$\displaystyle t_s(\lambda_s) = \frac{L}{v_g(\lambda_s)}$ (where $v_g$ is the group velocity inside the fiber spool) and that 
\begin{equation} \label{eq:t_sToD}
\begin{split}
d t_s
 = \frac{d \left(\frac{L}{v_g(\lambda_s)}\right)}{d\lambda_s} d\lambda_s 
                         = L D(\lambda_s) \ d\lambda_s,
\end{split}
\end{equation}
we have the equality:
\begin{equation}\label{nToNtilde}
	\tilde{n}(\lambda_s, \lambda_i) = L^2 D(\lambda_s)  D(\lambda_i)\	 {n}(t_s, t_i).
\end{equation}
In our case, this spectral-to-temporal mapping is one-to-one because the downconverted photons generated from the poled fiber 
have wavelengths that lie only in the anomalously 
dispersive region of the fiber spool used (SMF-28).  

Note that while $D(\lambda)$ is effectively the group-velocity dispersion (GVD) 
of the fiber spool, it encapsulates higher-order dispersion terms of the spool as well.  

\begin{figure}[htb]
	\centering
	\includegraphics[width=7cm]{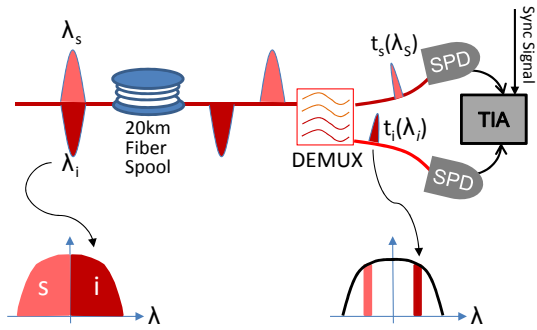}
	\caption{\label{fig:FiberSpectromPrelimsetup}
	A simple schematic of the fiber spectrometer setup.
	The spectral information of the 
	signal and idler photons are mapped onto the time domain 
	after traveling through the long fiber spool.  
	A WDM (labeled DEMUX) is used to route the shorter-wavelength signal to one 
	detector, and idler to the other detector, while also limiting the spectral extent 
	of each photon.  
	This is important, as it limits the temporal extent of the signal photon, which
	 cannot exceed 24.5 ns (the sync period of the Ti:Sapph oscillator).   
	}
\end{figure}
\begin{figure*}[htb]
\centering
\includegraphics[width=16cm]{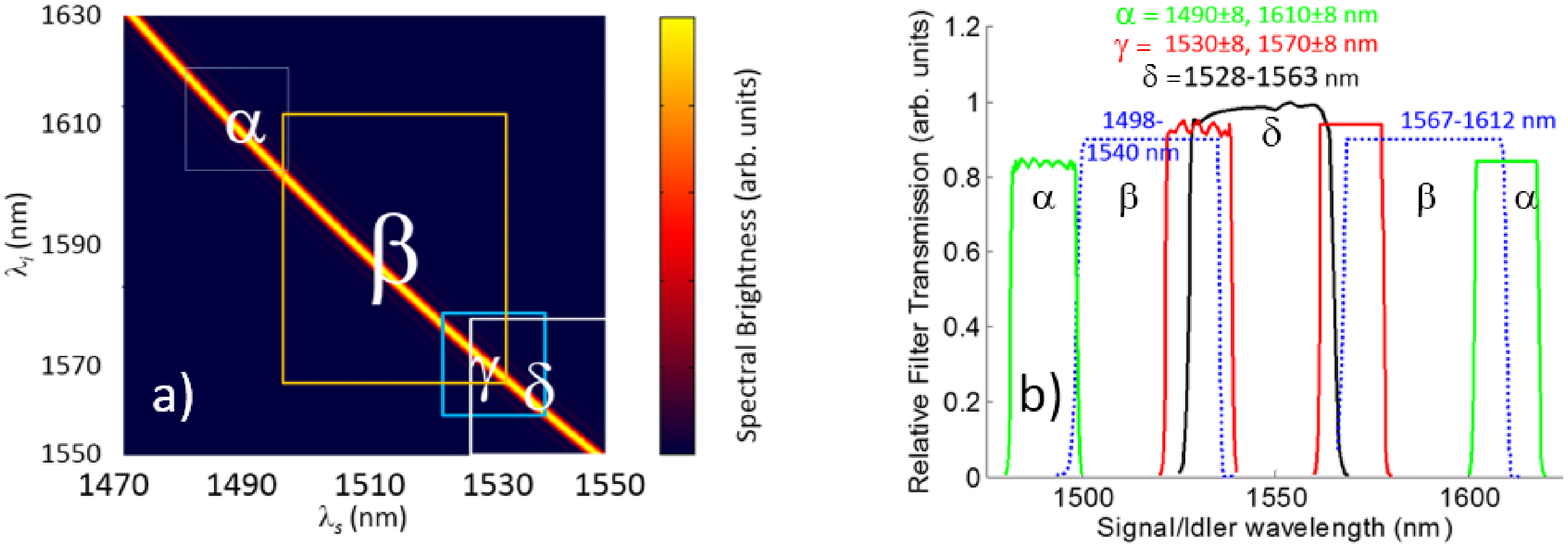}
\caption{\label{fig:JSA_FilterSets}
a) The various top-hat filters used are overlayed on top of the calculated joint spectra for 
the downconverted photon pairs.  
b) The (relative) transmission spectra of the filters used are plotted.  
$\alpha$) $\lambda_s = 1482-1498\text{ nm}$, $\lambda_i = 1602-1618 \text{ nm}$.
$\beta$) $\lambda_s = 1498-1540\text{ nm}$, $\lambda_i = 1567-1612 \text{ nm}$.
$\gamma$) $\lambda_s = 1522-1538\text{ nm}$, $\lambda_i = 1562-1578 \text{ nm}$.
$\delta$) The degenerate filter; the filter is from 1528 nm to 1563 nm. 
}
\end{figure*}
\begin{figure}[thb] 
\centering
\includegraphics[width=7cm]{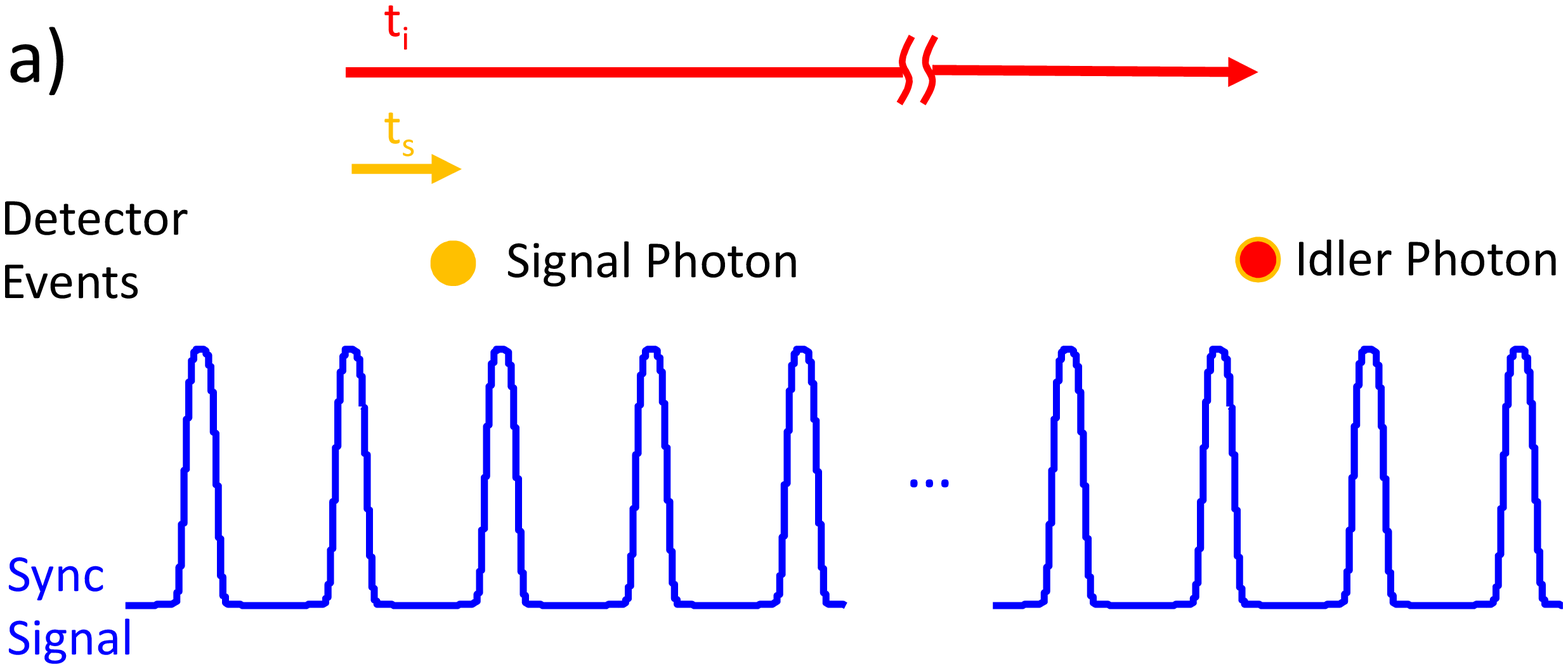}
\vspace{+10pt}
\includegraphics[width=7cm]{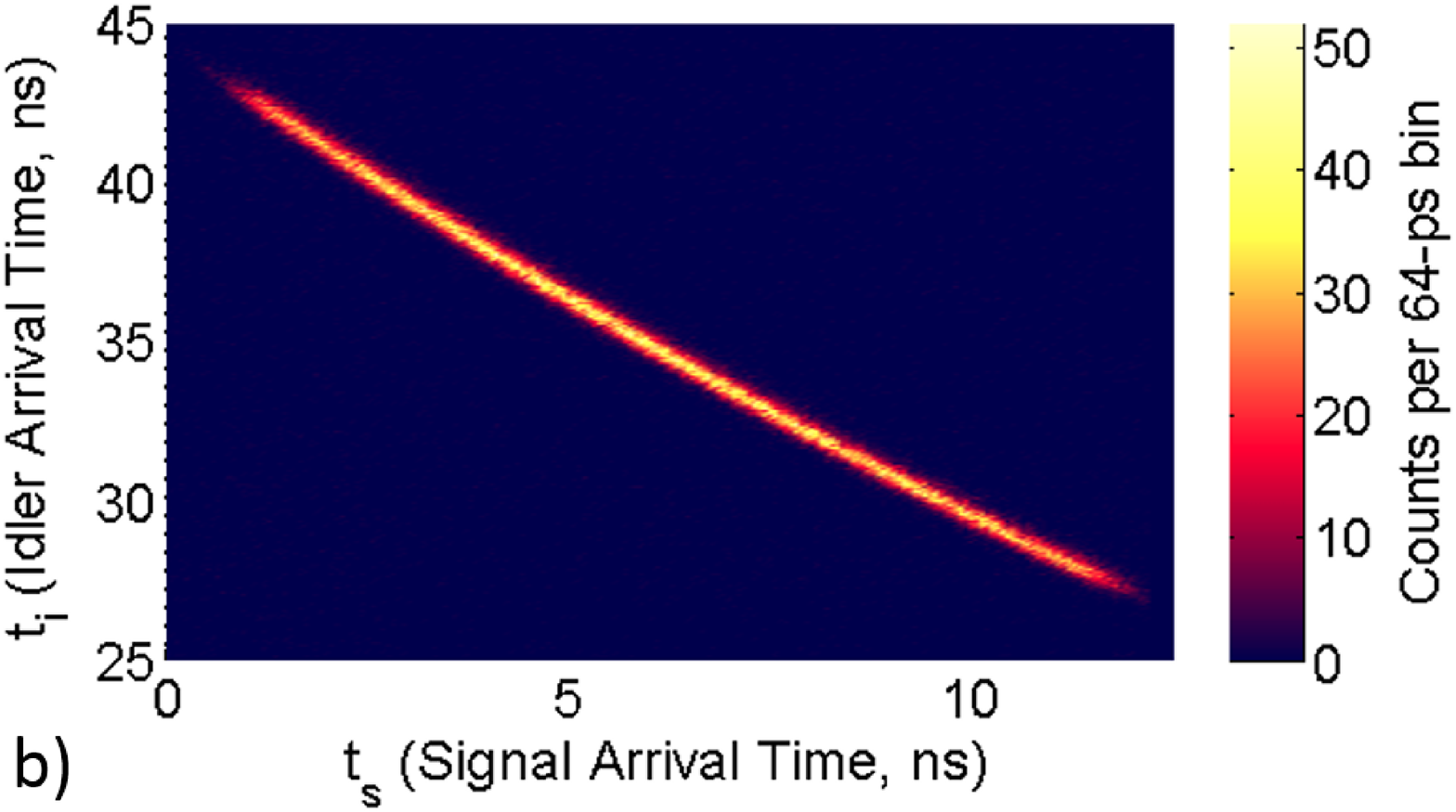}
\includegraphics[width=7cm]{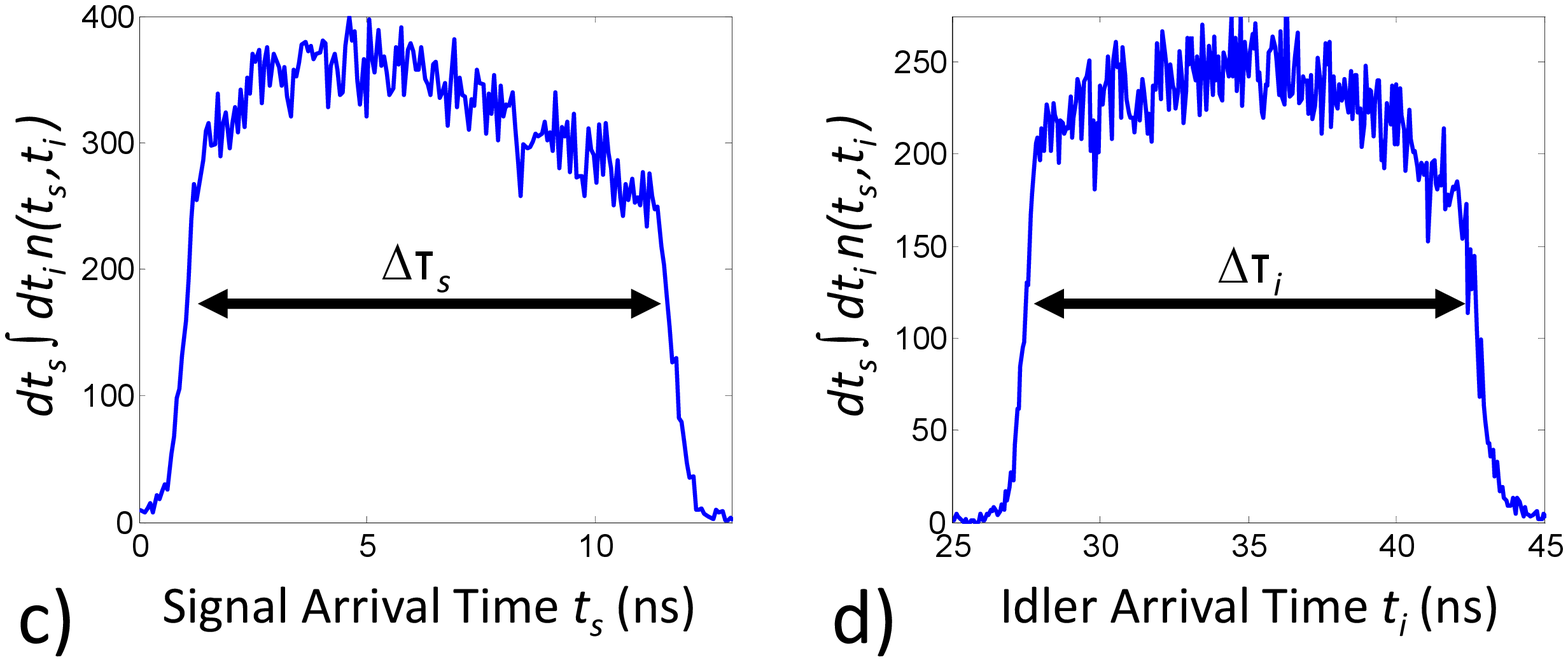}
\caption{\label{fig:SchematicT3Mode}
a) Schematic showing how ${n}(t_s,t_i)\ d t_s\ dt_i$ (Eqn. \ref{eq:biphotTime}) is generated from the time-interval anlyzer (TIA).  The
arrival times $t_s$ and $t_i$ are measured with respect to the Ti:Sapph sync signal, which has a 40.8 MHz repetition rate (24.5 ns period), and are discretized into ($dt_s = dt_i = $) 64-ps bins.  A two-dimensional histogram (b) is generated as the counts in each bin are allowed to accumulate over a long integration time, usually more than 500 seconds.  The frequency-conjugate filter sets used to 
generate this histogram were top-hat filters with bandwidths from 1498-1540 nm (signal) and 1567-1613 nm (idler).
(c)-(d) One-dimensional histograms are obtained when we integrate over the idler and signal arrival times, respectively.  The sharp rising and falling edges of these figures give us a
way to characterize the dispersion of our fiber spool, as we also know the (filter-limited) spectral extent of the biphotons that produce these histograms.   
}
\end{figure}

Experimentally, an actively-modelocked Ti:Sapphire laser 
with an 81.6 MHz pulsetrain and 400-ps FWHM pulses  
pumps a periodically-poled fiber to generate an ensemble of biphotons.  
Fig. \ref{fig:FiberSpectromPrelimsetup} shows an outline of the 
experimental setup.  
The signal photons (shorter wavelength) are separated from the idler photons (longer wavelength) using
coarse wavelength-division demultiplexers (DEMUX); the bandwidths of such coarse filters range 
from 16 nm to 40 nm (see Fig. \ref{fig:JSA_FilterSets}).  
Free-running {SPD}s, the id Quantique id220 devices, are used to
measure the photons.   
The arrival times of the signal and idler photons at the SPDs are measured with respect to the sync signal of the Ti:Sapph with the commercial time-interval analyzer (PicoQuant HydraHarp 400).
A two-dimensional histogram is constructed from the detector clicks and 
Ti:Sapph sync signal. 
A schematic for the process is shown in Fig. \ref{fig:SchematicT3Mode}.

Equation \ref{eq:t_sToD} allows us to relate the  temporal resolution to  the differential wavelength resolution $d\lambda_{\text{min}}$.  
The minimal resolution $dt_{\text{min}}$ is dictated by the jitter of the sync pulse ($\sigma_{\text{sync}}$), the pump pulsewidth ($\sigma_{\text{pulse}}$), as well as the jitter of the detectors ($\sigma_{\text{Det}}$).  We make the (good) approximation  that all three are normally-distributed $\left(\sim \exp\left(-\frac{t^2}{2\sigma^2}\right)\right)$, so that the jitters will add in quadrature to give the minimal resolution $dt_{\text{min}}$:
\begin{equation}\label{eq:GetMinRes}
	dt_{\text{min}} = \sqrt{
							\sigma^2_{\text{sync}} + \sigma^2_{\text{pulse}} + 
							\sigma^2_{\text{Det}}
							}.
\end{equation}
where $\sigma_{\text{sync}} < 30$ ps, 
$\sigma_{\text{pulse}}= 170$ ps (as $\text{FWHM} \approx 2.355\sigma$), 
and $\sigma_{\text{Det}} = 51$ ps.  
We see that the expression is dominated by $\sigma_{\text{pulse}}$.  The value of 
$dt_{min}$ (180 ps) corresponds to a 
wavelength resolution $d\lambda_\text{min}$ of 0.13 nm at 1550 nm for a 20-km spool of Corning SMF-28 fiber.  
%

Finite-bandwidth filters are required as the range of the arrival times of the signal photon must be no greater than 24.5 ns (or one period of the laser sync signal, 40.8 MHz).  
Multiple filter sets (Fig. \ref{fig:JSA_FilterSets})  
allow us to span a large
portion of the  downconversion spectrum (more than 100 nm).  
%
%

An average photon pair $\bar{N}_{ph}$ of no more than 0.01 pairs/pulse is generated over each filter set's bandwidth.  

Using Eq. \ref{nToNtilde}, we obtain the `raw' spectral brightness $\tilde{n}_{RAW}(\lambda_s, \lambda_i)$; we say it is raw because we still need to account for the spectral dependence of the insertion loss of the system (including detector efficiency and the system insertion loss), and normalize by the average pump power and detector deadtimes
in order to obtain the true spectrum of the biphoton.  
Finally, after obtaining $\tilde{n}(\lambda_s, \lambda_i)$, we can relate it to the biphoton wavefunction $\tilde{\phi}(\lambda_s, \lambda_i)$, where: 
\begin{equation}
|\psi\rangle = \int d\lambda_s\ d\lambda_i\ \tilde{\phi}(\lambda_s, \lambda_i),
\end{equation}
in the following way:
\begin{equation}
	\tilde{n}(\lambda_s, \lambda_i) = \hat{\mu}|\tilde{\phi}(\lambda_s, \lambda_i)|^2.
\end{equation}
The factor $\hat{\mu}$ is the  total brightness, defined 
as the average number of photon pairs generated (over the entire
downconversion spectrum) per mW of pump power.  

\afterpage{\clearpage}

\section{Fiber Dispersion Characterization}
\label{sec:FiberDispersionChar}
The dispersive medium used here for the spectrometer is a 20-km reel of SMF-28 fiber.
While the 
GVD of SMF-28 is well documented, there may be some 
manufacturing variability from one reel to another.  In this subsection, we detail how we measure the dispersion of a fiber spool used for the experiment.  But first, we must assume a form for the $D$ parameter \cite{agrawal2005lightwave}:
\begin{equation} \label{eq:DparamFit}
	D(\lambda) = \frac{S_0}{4}\left( \lambda - \frac{\lambda_0^4}{\lambda^3}\right)
\end{equation}
where $\lambda_0$ is the zero-dispersion wavelength, and $S_0$ is the dispersion slope at $\lambda_0$.  

We wish to know $S_0$ and $\lambda_0$ for our particular fiber spool, 
as well as the suitability of 
Eq. \ref{eq:DparamFit} over the wavelength range of interest (1480-1620 nm).  
To that end, we can exploit the filtered downconverted light  from our biphoton source to characterize the dispersion of the fiber spool.    
We observe that in Figs. \ref{fig:SchematicT3Mode}c and \ref{fig:SchematicT3Mode}d, the one-dimensional 
histograms of the signal and idler arrival times (respectively) have sharp rising and falling edges 
($> 15$ dB/nm).  These sharp features 
correspond to the cut-on ($\lambda_1$) and cut-off ($\lambda_2$) wavelengths of our filters, and using multiple filter sets (1490-1610, C-/L-Band filters, 1530-1570) to filter the downconverted light, we can perform a chi-square fit, using the formula below (Eqn. \ref{eq:FitForD}):

\begin{eqnarray}\label{eq:FitForD}
	\Delta\tau =
  L \int_{\lambda_1}^{\lambda_2} d\lambda
	          \ D\left(\lambda
	           \right),
\end{eqnarray}

\noindent
with the cut-on/off wavelengths as the independent variable, and the range $\Delta\tau$ of arrival times as the dependent variable.
The best fit yields the following results: 
$S_0 = 0.0885\pm 0.0006$ ps/(nm$^2\cdot$km), $\lambda_0 = 1320\pm 1.7$ nm, and $L = 20.56 \pm 0.13$ km.

With the fiber spool dispersion characterized, we are ready to use it to invert the temporal histogram data $n(t_s,t_i)$ to obtain the biphoton spectral density 
$\tilde{n}(\lambda_s,\lambda_i )$.

\afterpage{\clearpage}

\section{Experiment and Results}
\label{sec:FibSpecExptResults}

The fiber spectrometer method is used in the measurement of  the joint spectrum of a broadband correlated photon pair source, namely the poled optical fiber \cite{bonfrate1999parametric,PhanHuy:07,zhueyiPRL2012, zhueyiOL2013}.  
The poled fiber has a non-zero second-order nonlinearity, and is quasi-phase-matched \cite{corbari2005aff} for the downconversion of biphotons at $\sim 1.55\ \mu{\text{m}}$.
The type-II downconversion process \cite{zhueyi2010OL,TwistPaperJOSAB2010} is exploited; however, neither 
the polarization-entangled nature \cite{helt2009proposal,zhueyiPRL2012, zhueyiOL2013} nor the polarization degree-of-freedom in general are exploited.  


The source's phase-matching peak occurs at the pump wavelength  $\lambda_{p0} = $ 774.5 nm, with a FWHM pump acceptance bandwidth of 0.35 nm.
The pump pulsetrain used is narrowband ($< 0.02$ nm) compared to this bandwidth.  
Due to the broadband nature of the downconverted photons, and the narrowband pump used, the generated biphotons are highly spectrally-correlated (Fig. \ref{fig:JSA_FilterSets}a).  
Additionally, the biphoton spectrum varies significantly as a function of pump wavelength.  

\begin{figure}[htb]
	\includegraphics[width=8cm]{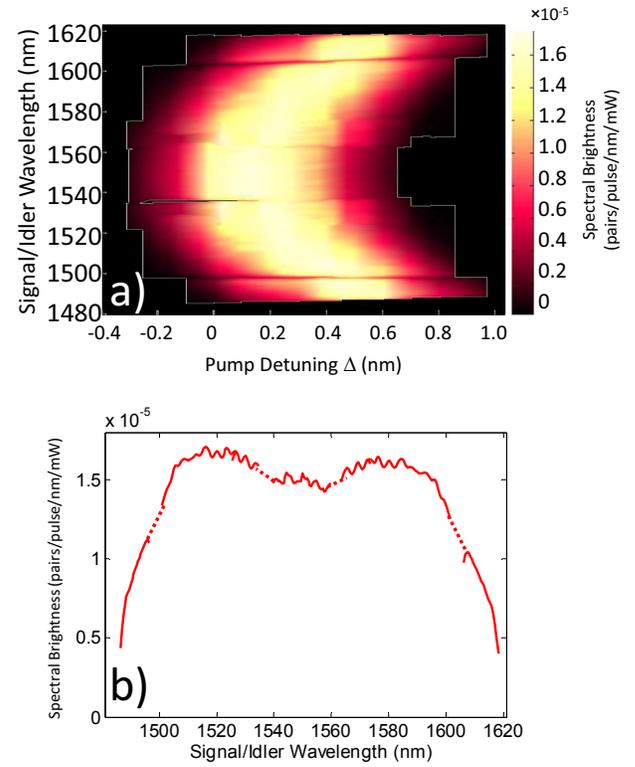}
	\caption{\label{fig:results}
	a)  The tuning curve (spectral brightness plotted as a function of signal/idler and pump wavelength) is obtained for the downconverted photon pairs generated by the poled optical fiber.   
	b)  A cross-section of the tuning curve is shown  where the downconversion 
	bandwidth is largest; this corresponds to a pump detuning $\Delta = 0.25$ nm.  
	Dotted lines are used to fill in the gaps in the spectrum where no data is available.  
	}
\end{figure}

The experimental procedure is as follows.  The pump   is set to a particular wavelength
 $\lambda_p$, and generates downconverted biphotons when launched into the source.  The biphotons are then coupled into the dispersive fiber spool 
(Fig. \ref{fig:FiberSpectromPrelimsetup}), and then filtered with frequency-conjugate filters 
(Fig. \ref{fig:JSA_FilterSets}) at the output.  A 2-D temporal histogram ($n(t_s,t_i)$)
tallying up the coincidences as a function of signal and idler photon arrival time 
is collected (Fig. \ref{fig:SchematicT3Mode}b).  Integration times of 500-1000 seconds are used 
to generate each histogram.  
The temporal extent for the 2-D histogram is
measured along the $t_s$ ($\Delta\tau_s$, Fig. \ref{fig:SchematicT3Mode}c) 
and $t_i$ axes ($\Delta\tau_i$, Fig. \ref{fig:SchematicT3Mode}d), as these values are used to obtain 
the fiber spool's dispersion characteristics (Eqn. \ref{eq:FitForD}).
The procedure is repeated with another pair of filters or pump wavelength until a 120 nm range at the signal/idler wavelength and a 1 nm range 
about the central pump wavelength $\lambda_{p0}$ is swept (see Fig. \ref{fig:results}a).  The pump wavelength is
varied at a step size of approximately 0.1 nm.    

At certain pump wavelengths, not every filter pair will be completely frequency-conjugate.  
For example, consider the C-Band (L-Band) filter, that has cut-on and cut-off wavelength 
from 1498 to 1540 nm (1567.6 to 1612.6 nm).  Table \ref{tab:FibCalib} gives the 
signal filter's actual  frequency-conjugate bandwidths (from $\lambda_1$ to $\lambda_2$) and temporal extent $\Delta\tau_s$ (Fig. \ref{fig:SchematicT3Mode}c) 
at various pump wavelengths.

\begin{table}[h] 
  \begin{center}
    \begin{tabular}{c c c c }
        \hline	
        \hline
         $\lambda_p$ & $\lambda_{1}$  & $\lambda_{2}$  & Time-Width \\
(nm) &         (nm)                                       &      (nm)                                      & $\Delta\tau_s$ (ns)\\
        \hline
        \hline
	774.7 	 & 1498      &  1531.7    &  9.77 \\
	774.9 	& 1498     &   1532.5    & 10.05 \\
	775.1 	& 1498    &   1533.1    & 10.20 \\
    \hline	
    \end{tabular}
	\end{center}
	\caption{\label{tab:FibCalib}
Data obtained using the C-Band filter as a function of pump wavelength.  
The cut-on and cut-off wavelengths refer to the portion of the filter where 
photons that have conjugates residing in the transmission window of the 
L-Band filter are generated.  
The overall bandwidth for the C-Band (L-Band filter) 
is 1498-1540 nm (1567.6-1612.6 nm).
The error for $\Delta\tau_s$ is taken to be $dt_{min}$ (see Eqn. \ref{eq:GetMinRes}), or
$\pm180$ ps.
	}
\end{table}

Data similar to Table \ref{tab:FibCalib} is collected from all the experimental runs for various
pump wavelengths and filter sets, and fed into Eqns. \ref{eq:DparamFit} and \ref{eq:FitForD}, from which the dispersion properties ($D(\lambda)$) 
of the fiber spool are then determined (as described in Section 3).

Once the $D-$parameter of the dispersive spool is known, the 2-D temporal histogram 
$n(t_s,t_i)$ (see Fig. \ref{fig:SchematicT3Mode}b) can then be related via 
Eqn. \ref{nToNtilde} to $\tilde{n}(\lambda_s,\lambda_i)$.  The data for the various 
filter sets and pump wavelengths 
is then stitched together 
 to give a tuning curve (Fig. \ref{fig:results}a).  The $x-$axis is the pump detuning $\Delta$ ($\Delta \equiv \lambda_{p}-\lambda_{p0}$), while the $y-$axis is the signal/idler wavelength.
 A cross-section of this tuning curve (at $\Delta = 0.25$  nm) is shown in  
Fig. \ref{fig:results}b.

%
%

\vspace{10pt}
In summary, we have demonstrated a method for measuring the broadband 
downconversion spectrum of a correlated photon pair source using a fiber
spectrometer.  The dispersion of the fiber spool can also be simultaneously measured using the same source of broadband correlated photon pairs.

\section{Acknowledgements}

PGK and CC acknowledge the EU Project CHARMING 
(Contract No. FP7-288786) for financial support.  
LQ and EYZ wish to acknowledge NSERC (the Natural Sciences and Engineering Research Council of Canada), CFI (Canada Foundation for Innovation), and 
ORF (Ontario Research Fund) for funding the work presented in this paper.



\end{document}